\documentclass[prb,twocolumn,showpacs,amsmath,amssymb]{revtex4}
\usepackage{graphicx} % include fig
\usepackage{dcolumn} % align table by decimal point

\begin{document}
\title{Singlet-triplet splitting, correlation and entanglement of
two electrons in quantum dot molecules}

\author{Lixin He}
%\affiliation{National Renewable Energy Laboratory, Golden, Colorado 80401}
\author{Gabriel Bester}
%\affiliation{National Renewable Energy Laboratory, Golden, Colorado 80401}
\author{Alex Zunger}
\affiliation{National Renewable Energy Laboratory, Golden, Colorado 80401}
%\email{lhe@nrel.gov}
%\homepage{sst.nrel.gov}

\date{\today}
\begin{abstract}

Starting with an accurate pseudopotential description of the single-particle
states, 
and following by configuration-interaction
treatment of correlated electrons in vertically coupled,
self-assembled InAs/GaAs quantum dot-molecules, we show how simpler,
popularly-practiced approximations,
depict the basic physical characteristics including the singlet-triplet
splitting, degree of entanglement (DOE) and correlation. The mean-field-like
single-configuration approaches such as Hartree-Fock and local spin
density, lacking correlation, incorrectly identify the ground state symmetry
and give inaccurate values for the singlet-triplet splitting and the DOE. 
The Hubbard model gives qualitatively correct
results for the ground state symmetry and singlet-triplet splitting,
but produces significant errors in the DOE because it ignores the fact that
the strain is asymmetric even if the dots within a molecule are identical.
Finally, the Heisenberg model gives qualitatively correct ground state
symmetry and singlet-triplet splitting only for rather large inter-dot
separations, but it greatly overestimates the DOE
as a consequence of ignoring the electron double occupancy effect.

\end{abstract}
\maketitle

\section{Introduction}  

\begin{figure}
\includegraphics[width=3.5in,angle=0]{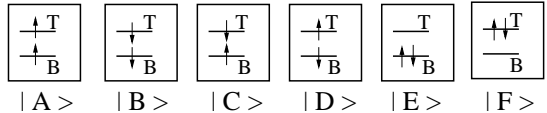}
\caption{Possible configurations for two electrons in two vertically coupled
  quantum dots. (a) Spin configurations in the MO basis. $\sigma_g$ and
  $\sigma_u$ indicate the bonding and anti-bonding states respectively.
(b) Spin configurations in the dot-localized basis.  
``T'' and ``B'' indicate the top and bottom dot. 
}
\label{fig:CI_basis}
\end{figure}

Two vertically\cite{pi01, rontani04} or laterally\cite{waugh95} coupled 
quantum dots containing electrons, holes, or an exciton constitute  
the simplest solid structure proposed for the basic 
gate operations of quantum computing.\cite{bayer01,loss98}
The operating principle is as follows:
when two dots couple to each other, bonding and anti-bonding
``molecular orbitals'' (MO) ensue from the single-dot orbitals
\{$\chi_i$\} of the top (T) and bottom (B) dots:
$\psi(\sigma_g)=\chi_T(s)+\chi_B(s)$ is the $\sigma$-type bonding 
and $\psi(\sigma_u)=\chi_T(s)-\chi_B(s)$ is the $\sigma$-type antibonding
state. 
Similarly,   $\psi(\pi_u)=\chi_T(p)+\chi_B(p)$ 
and $\psi(\pi_g)=\chi_T(p)-\chi_B(p)$ are the ``$\pi$'' 
bonding and anti-bonding states constructed 
from the ``p'' single-dot orbitals of top and
bottom dots, respectively. 
Injection of two electrons into such a diatomic ``dot-molecule'' creates 
different spin configurations such as 
$|\sigma_{g}^{\uparrow},\sigma_{u}^{\uparrow}\rangle$ or
$|\sigma_{g}^{\downarrow},\sigma_{u}^{\downarrow}\rangle$,
depicted in Fig.~\ref{fig:CI_basis}a.
In the absence of spin-orbit coupling, these two-electron states are either
spin-singlet or spin-triplet states with energy separation $J_{S-T}$.
Loss and DiVincenzo\cite{loss98} proposed a ``swap gate'' base on 
a simplified model, where two localized spins have Heisenberg coupling,
$H=J_{S-T}(t)\vec{S}_1\cdot\vec{S}_2$. Here $\vec{S}_1$ and 
$\vec{S}_2$ are the spin-${1\over2}$ operators for the two localized
electrons. The effective Heisenberg exchange splitting $J_{S-T}(t)$ is
a function of time $t$, which is measured as the difference in the
energy between the spin-triplet state with the total spin $S=1$ and the
spin-singlet state with $S=0$. 
The ``state swap time'' is $\tau \sim 1/J_{S-T}$.
An accurate treatment of the singlet-triplet splitting
$J_{S-T}$ and the degree of entanglement carried by the
two electrons is thus of outmost importance for 
this proposed approach to quantum computations.
 
Theoretical models, however, differ in their assessment of
the magnitude and even the sign of the singlet-triplet energy difference
$J_{S-T}$ that can be realized in a quantum dot
molecule (QDM) with two electrons. Most theories have attempted to
model dot molecules made of large (50 - 100 nm),
electrostatically-confined\cite{ashoori92,johnson92,tarucha96} 
dots having typical 
single-particle electronic levels separation of 1 - 5 meV, 
with larger (or comparable) inter-electronic
Coulomb energies $J_{ee} \sim$ 5 meV.
The central approximation used almost universally is that the single-particle
physics is treated via particle-in-a-box
effective-mass approximation (EMA), where multi-band and
intervally couplings are neglected. 
(In this work, we will deviate from this tradition, see below)
Many-body treatments of this simplified EMA
model range from phenomenological Hubbard\cite{burkard99} or Heisenberg
\cite{loss98,burkard99} 
models using empirical input parameters, to microscopic 
Hartree-Fock (HF)~\cite{tamura98,yannouleas99,hu00},
local spin densities (LSD) approximation~\cite{nagaraja99,partoens00} 
and configuration
interaction (CI) method.~\cite{rontani01,rontani04}

The LSD-EMA~\cite{nagaraja99,partoens00} can treat easily up to 
a  few tens of electrons in the quantum dot molecules, 
but has shortcoming for treating strongly correlated 
electrons, predicting for a dot molecule loaded with two electrons that
the triplet state is  {\it below} the singlet in the weak coupling region,
\cite{nagaraja99} 
as well as incorrectly
mixing singlet (spin unpolarized) and triplet (spin polarized) even in the
absence of spin-orbit coupling.
Since in mean-field approaches like LSD or HF, the
two electrons are forced to occupy the same
molecular orbital delocalized on both dots, the two-electron states
are purely unentangled.

The Restricted (R)HF method (RHF-EMA)  shares similar failures with LSD, 
giving a triplet as the
ground state at large inter-dot separation.
The Unrestricted (U) HF~\cite{yannouleas99}
corrects some of the problems of RHF by 
relaxing the requirement of
(i) two electrons of different spins occupying the same spatial orbital,
and (ii) the single-particle wavefunctions have the symmetry of the external
confining potential.  
The UHF-EMA correctly give the singlet lower in energy than the triplet,
\cite{hu00} 
and can also predict Mott localization of the electrons in the
dot-molecule, which
breaks the many-particle symmetry.~\cite{yannouleas99} 
However, since in UHF, the symmetry-broken wavefunctions are 
only the eigenstates
of the $z$-component of total spin $S=s_1+s_2$, but not of $S^2$,  
the UHF-EMA incorrectly mixes the singlet 
and triplet.~\cite{hu00,yannouleas99} 
For the simple case of dot molecules having inversion symmetry, 
(e.g. molecules made of spherical dots but not of vertical lens-shaped
dots), assuming EMA and neglecting spin-orbit coupling, 
there is an {\it exact} symmetry.  
For this case, Ref. \onlinecite{yannouleas01, yannouleas02} 
indeed were able to project out 
the eigenstates of $S^2$,  yielding good spin quantum numbers
and lower energy. 
However, for vertically coupled lens shaped quantum dots (i.e., realistic
self-assembled systems)
or even for spherical dots, but in the presence of spin-orbit coupling, 
there is no exact symmetry.
In this case,
configurations with
different symmetries may couple to each other.
To get the correct energy spectrum
and many-body wavefunctions, 
a further
variation has to be done after the projection, e.g. using the 
Generalized Valence Bond (GVB) method.\cite{goddard73}
For this case and other cases a CI approach is needed.
 
The CI-EMA has been proven\cite{rontani01,rontani04} to be 
accurate for treating
few-electron states in large electrostatic dot molecules, and  
predicts the correct ground state.
Finally, recent Quantum Monte Carlo-EMA calculations~\cite{das_unpub}
also show that the singlet is below the triplet.

The above discussion pertained to large (50 - 100 nm) 
electrostatic-confined dots.  
Recently, dot molecules have been fabricated\cite{ota03,ota04a} 
from self-assembled
InAs/GaAs, offering a much larger $J_{S-T}$. Such dot have much smaller
confining dimensions (height of only 2 - 5 nm), showing a
typical spacing between
electron levels of 40 - 60 meV, {\it smaller} interelectronic Coulomb energies
$J_{ee}\sim$ 20 meV and exchange energies of $K_{ee}\sim$ 3  meV. 
Such single 
dots have been accurately modeled\cite{wang99c} 
via atomistic pseudopotential theories,
applied to the single-particle problem (including multi-band and intervally
couplings as well as non-parabolicity, 
thus completely avoiding the effective mass
approximation). The many-particle problem is then described
via all-bound-state configuration-interaction method. 
Here we use this methodology
to study the singlet-triplet splitting in vertically-stacked self-assembled
InAs/GaAs dots. We calculate first 
the singlet-triplet 
splitting {\it vs} inter-dot separation, finding the singlet to be
below the triplet.
We then simplify our model in successive steps, reducing the sophistication
with which interelectronic correlation is described and showing how these
previously practiced
approximations~\cite{tamura98,yannouleas99,hu00,nagaraja99,partoens00} 
lead to different values of $J_{S-T}$,
including its sign reversal. This methodology provides insight into the
electronic processes which control the singlet-triplet splitting in
dot-molecules.
 
The remainder of the paper is arranged as follows. In Sec.~\ref{sec:methods}
we provide technical details regarding
the methodology we use for the calculations. 
We then compare the singlet-triplet splitting, degree of entanglement
and correlation of two-electron states in 
different levels of approximations in Sec.~\ref{sec:results}. 
Finally, we summarize in Sec.~\ref{sec:summary}.

\section{methods}
\label{sec:methods}

\subsection{Geometry and strain relaxation}

\begin{figure}
\includegraphics[width=3.in,angle=0]{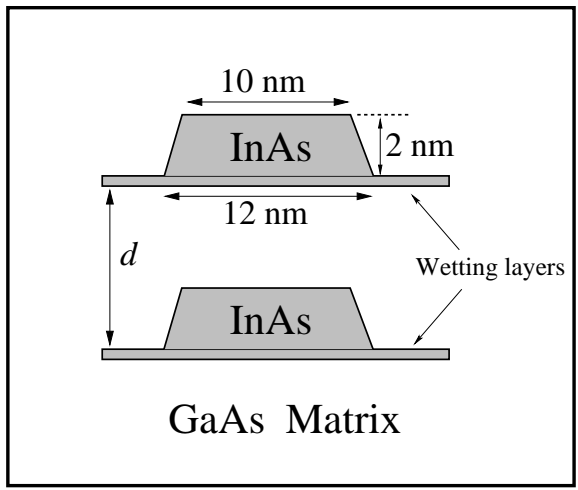}
\caption{
Geometry of the two vertically coupled quantum dot molecule.
The inter-dot distance $d$ is measured from wetting layer to wetting layer.}
\label{fig:geom}
\end{figure}

We consider a realistic dot-molecule geometry\cite{bayer01}
shown in Fig.~\ref{fig:geom}, which has 
recently been used in studying exciton entanglement,
\cite{bayer01,bester04b}
and two-electron states.\cite{he05b}
Each InAs dot is 12 nm wide and 2 nm tall, with one monolayer 
InAs ``wetting layer'', and compressively strained
by a GaAs matrix. 
Even though experimentally grown dot molecules often have slightly different
size and composition profile for each dot within the molecule, 
here we prefer to consider
identical dots, so as to investigate the extent of symmetry-breaking due to
many-body effects in the extreme case of identical dots.
The minimum-strain configuration is
achieved at each inter-dot separation $d$, by relaxing
the positions $\{{\bf R}_{n,\alpha}\}$ of all (dot + matrix) 
atoms of type $\alpha$ at site $n$, so as to
minimize the bond-bending and
bond-stretching energy using the Valence Force Field (VFF)
method.~\cite{keating66,martins84}
This shows that both dots have large and
nearly constant hydrostatic strain inside the dots which decays rapidly 
outside.\cite{he05b} 
However, even though the dots comprising the molecule are geometrically
identical,
the strain on the two dots is different 
since the molecule lacks inversion symmetry. In fact, 
we found that
the top dot is slightly more strained than the bottom dot. 
Not surprisingly, the GaAs region {\it between} the two dots is
more severely strained than in other parts of the matrix, as shown
in Fig. 1 of
Ref.~\onlinecite{he05b} and as the two dots move apart,
the strain between them decreases.

\subsection{Calculating the single-particle states}

The single-particle electronic energy levels and wavefunctions 
are obtained by solving the
Schr\"{o}dinger equations in a pseudopotential scheme,
\begin{equation}
\left[ -{1 \over 2} \nabla^2 
+ V_{\rm ps}({\bf r}) \right] \psi_i({\bf r})
=\epsilon_i \;\psi_i({\bf r}) \; ,
\label{eq:schrodinger}
\end{equation}
where the total electron-ion potential $ V_{\rm ps}({\bf r})$ 
is a superposition of
local, screened atomic pseudopotentials $v_{\alpha}({\bf r})$, i.e.
$V_{\rm ps}({\bf r}) =\sum_{n,\alpha} 
v_{\alpha}({\bf r} - {\bf R}_{n,\alpha})$.
The pseudopotentials used for InAs/GaAs are identical to those
used in Ref.~\onlinecite{williamson00} and were tested for 
different systems.\cite{williamson00,bester04b, he04a} 
We ignored spin-orbit coupling in the InAs/GaAs quantum dots,
since it is extremely small for electrons treated here (but not for holes
which we do not discuss in the present work).  
Without spin-orbit coupling, the states of two electrons are
either pure singlet, or pure triplet. 
However, if a spin-orbit coupling is introduced, the singlet
state would mix with triplet state.

Equation~(\ref{eq:schrodinger}) is solved using the ``linear 
combination of Bloch
bands'' (LCBB) method,\cite{wang99b}
where the wavefunctions $\psi_i$ are expanded as, 
\begin{equation}
\psi_{i}({\bf r}) =\sum_{n,{\bf k}}\sum_{\lambda}
C_{n,{\bf k}}^{(\lambda)}\;
\phi_{n,{\bf k},\tensor{\epsilon}}^{(\lambda)}({\bf r})\; .
\label{eq:lcbb}
\end{equation}
In the above equation,
$\{\phi_{n,k,\tensor{\epsilon}}^{(\lambda)}({\bf r})\}$ are the bulk
Bloch orbitals of band index $n$
and wave vector ${\bf k}$ of material $\lambda$ (= InAs, GaAs), strained
uniformly to strain $\tensor{\epsilon}$. The dependence of the basis functions
on strain makes them variationally efficient.
(Note that the potential $V_{\rm ps} ({\bf r})$ itself also has 
the inhomogeneous strain dependence through the atomic position 
${\bf R}_{n,\alpha}$.)
We use  for the basis set $\tensor{\epsilon}=0$ for the
(unstrained) GaAs matrix material, and an average $\tensor{\epsilon}$ value 
from VFF for the strained dot material (InAs).
For the InAs/GaAs system, we use $n=2$ (including spin) 
for electron states on a
6$\times$6$\times$28 k-mesh.
A single dot with the geometry of Fig.\ref{fig:geom} (base=12 nm and
height=2 nm) has three bound electron
states ($s$, $p_1$, and $p_2$) and more than 10 bound hole states. 
The lowest exciton transition in the single dot occurs
at energy 1.09 eV.
For the dot molecule
the resulting single-particle states are, in order of increasing energy, the
singly degenerated $\sigma_g$ and $\sigma_u$, (bonding and antibonding
combination of the s-like single-dot orbitals), 
and the doubly (nearly) degenerated 
$\pi_u$ and $\pi_g$, originating from doubly (nearly)
degenerate ``p'' orbitals (split by a few meV) in a
single dot.
Here, we use the symbols $g$ and $u$ to denote symmetric and anti-symmetric
states, even though in our case the single-particle wavefunction are 
actually asymmetric.\cite{he05b}
We define
the difference between the respective dot molecule
eigenvalues as $\Delta_{\sigma}=\epsilon(\sigma_u)-\epsilon(\sigma_g)$ and 
$\Delta_{\pi}=\epsilon(\pi_g)-\epsilon(\pi_u)$.

\subsection{Calculating the many-particle states}

The Hamiltonian of interacting electrons
can be written as,
\begin{equation}
H=\sum_{i\sigma} \epsilon_{i}\psi^{\dag}_{i\sigma} 
\psi_{i\sigma} 
+ {1\over2} \sum_{ijkl} \sum_{\sigma,\sigma'}
\Gamma^{ij}_{kl}\;
\psi^{\dag}_{i\sigma}\, \psi^{\dag}_{j\sigma'}\,
\psi_{k\sigma'}\,\psi_{l\sigma}\, ,
\label{eq:morbit_h}
\end{equation}
where,  $\psi_{i}$= $\sigma_u$, $\sigma_g$, 
$\pi_u$, $\pi_g$ 
are the single-particle energy levels of the 
{\it $i$-th molecular orbital},
while $\sigma$, $\sigma'$=1, 2 are spin indices.
The $\Gamma^{ij}_{kl}$ are 
the Coulomb integrals between molecular orbitals $\psi_{i}$, 
$\psi_{j}$,
$\psi_{k}$ and $\psi_{l}$,
\begin{equation}
\Gamma^{ij}_{kl} =\int\int d{\bf r}d{\bf r'}\; 
{\psi^*_{i}({\bf r}) \psi^*_{j} ({\bf r'}) 
\psi_{k}({\bf r'}) \psi_{l} ({\bf r}) 
\over \epsilon({\bf r}- \bf{r'}) |{\bf r} -{\bf r'}|} \, .
\label{eq:int_m}
\end{equation}
The $J_{ij}=\Gamma^{ij}_{ji}$ and
$K_{ij}=\Gamma^{ij}_{ij}$
are diagonal Coulomb and exchange integrals respectively.
The remaining terms are called off-diagonal or scattering terms.
All Coulomb integrals are calculated numerically from atomistic 
wavefunctions.~\cite{franceschetti99}
We use a phenomenological, position-dependent dielectric  
function $\epsilon({\bf r}- \bf{r'})$ to screen the electron-electron
interaction.\cite{franceschetti99}

We solve the many-body problem of Eq.(\ref{eq:morbit_h}) via the
CI method, by expanding the $N$-electron wavefunction
in a set of Slater determinants,
$|\Phi_{e_1,e_2,\cdots,e_N}\rangle
=\phi^{\dag}_{e_1}\phi^{\dag}_{e_2}\cdots\phi^{\dag}_{e_N}|\Phi_0\rangle$,
where $\phi^{\dag}_{e_i}$
creates an electron in the state $e_i$ .
The $\nu$-th many-particle wavefunction is then the linear combinations of
the determinants, 
\begin{equation}
|\Psi_{\nu}\rangle=\sum_{e_1,e_2,\cdots,e_N}
A_{\nu}(e_1,e_2,\cdots,e_N)\;|\Phi_{e_1,e_2,\cdots,e_N}\rangle \; .
\label{eq:coeff}
\end{equation}
In this paper, we only discuss the two-electron problem, i.e. $N$=2.
Our calculations include all possible Slater 
determinants for the six single-particle levels.

\subsection{Calculating pair correlation functions and degree of
entanglement}
\label{sec:p-corr}

We calculate in addition to the energy spectrum and the
singlet-triplet splitting $J_{S-T}$ also
the pair correlation functions and 
the degrees of entanglement (DOE).
The pair correlation function $P_{\nu}({\bf r}, {\bf r'})$ for an $N$-particle
system is defined as the
probability of finding an electron at ${\bf r'}$, given that 
the other electron is at ${\bf r}$, i.e.,  
\begin{equation}
P_{\nu}({\bf r}, {\bf r'})= \int d{\bf r}_3 \cdots d{\bf r}_N  |
 \Psi_{\nu}({\bf r},{\bf r'}, {\bf r}_3, \cdots,{\bf r}_N )|^2\; , 
\label{eq:corrfunc}
\end{equation}
where, $\Psi_{\nu}({\bf r}_1, \cdots, {\bf r}_N)$ is the $N$-particle
wavefunction of state $\nu$. For two electrons, the pair correlation function
is just $P_{\nu}({\bf r}, {\bf r'})= | \Psi_{\nu}({\bf r},{\bf r'})|^2$.

The degree of entanglement (DOE) is one of the most 
important quantities for
successful quantum gate operations. 
For {\it distinguishable} particles such as electron and hole, the DOE can be
calculated from Von Neumann-entropy formulation.
\cite{nielsen_book,bennett96,bennett96b,wehrl78} 
However, for {\it indistinguishable} particles, there are some
subtleties \cite{schliemann01a,paskauskas01,li01,zanardi02,
shi03,wiseman03,ghirardi04}
for defining the DOE 
since it is impossible to separate 
the two {\it identical} particles. 
Recently, a quantum correlation function
\cite{schliemann01a} has been proposed for 
indistinguishable particles using the Slater decompositions.
\cite{yang62}
We adapt this quantum correlation function 
to define the DOE for indistinguishable
fermions as,
\begin{equation}
\mathcal{S}=-\sum_{u} z_{i}^2 \;{\rm log}_2\, z_{i}^2\; ,
\label{eq:my-entropy-t}
\end{equation}
where, $z_i$ are Slater decomposition coefficients.
The details of deriving Eq. (\ref{eq:my-entropy-t}) are given is 
Appendix~\ref{sec:append-b}.
We also show in Appendix~\ref{sec:append-b} that 
the DOE measure Eq.(\ref{eq:my-entropy-t}) reduces to the usual 
Von Neumann-entropy formulation when the two-electrons
are far from each other.

\section{results}
\label{sec:results}

Figure~\ref{fig:t-j} shows the bonding-antibonding splitting
$\Delta_{\sigma}(d)$ between the molecular orbitals {\it vs} 
inter-dot separation $d$ measured from one wetting layer to the other,
showing also the value 
$\delta_{sp} =\epsilon_p-\epsilon_s$ of the
splitting between the p and s orbital energies of a {\it single dot} 
(i.e. $d\rightarrow \infty$). The bonding-antibonding splitting 
decays approximately exponentially as
$\Delta_{\sigma}=2.87\exp(-d/1.15)$ eV between $d \sim$4 - 8 nm.
The result of bonding-antibonding splitting includes 
two competing effects. On one hand, large interdot distance $d$
reduces the coupling between the two dots;
on the other hand, the strain between the dots is also
reduced, leading to a lower tunneling barrier, thus increases coupling. 
The local maximum of $\Delta_{\sigma}$ 
at $d$=8.5 nm is a consequence of the this competition.
Recent experiments~\cite{ota03,ota04a} 
show the  bonding-antibonding splitting of about 4 meV 
at $d$=11.5 nm for vertically coupled InAs/GaAs quantum dots molecules,
of similar magnitude as the value obtained here($\sim$ 1 meV),
considering that the measured dot molecule is larger (height/base= 4 nm/40 nm 
rather than 2 nm/12 nm in our calculations) and possibly asymmetric.
We also give in Fig.~\ref{fig:t-j} 
the interelectronic Coulomb energy $J_{C}$ of a {\it single-dot} s orbital. 
We define {\it strong coupling} region as 
$\Delta_{\sigma}$ $\approx$ $\delta_{sp}$, and
{\it weak coupling} region  $\Delta_{\sigma}$ $\ll $ $\delta_{sp}$. We see 
in Fig.~\ref{fig:t-j} strong
coupling for $d \leq $ 4 nm, and weak coupling for $d \geq $ 5 nm.
In the weak coupling region, 
the $\pi$ levels are well above the $\sigma$ levels. 
We also define ``strong confinement'' as $\delta_{sp} > J_{C}$, 
and weak confinement
as the reverse inequality. Figure~\ref{fig:t-j} shows that our dot is in the
strong-confinement regime.
In contrast, electrostatic dot
\cite{ashoori92,johnson92,tarucha96} are in the
weak confinement regime.

We next discuss the two-electron states in the QDMs and examine
several different approximations which we call Levels 1 - 4,
by comparing the properties of the ground states,
the singlet-triplet energy separation $J_{S-T}$ and 
the pair correlation
functions as well as the degree of entanglement for each state.
Starting from our most complete model (Level 1) 
and  simplifying it in successive steps, 
we reduce the sophistication
with which interelectronic correlation is described and show how these
previously practiced approximations lead to different values of $J_{S-T}$
(including its sign reversal), 
and different degree of entanglement. 
This methodology provides insight into the
electronic features which control singlet-triplet splitting and
electron-electron entanglement in dot molecules.

\subsection{Level-1 theory: all-bound-state configuration interaction}
\label{sec:level-1}

\begin{figure}
\includegraphics[width=3.5in,angle=0]{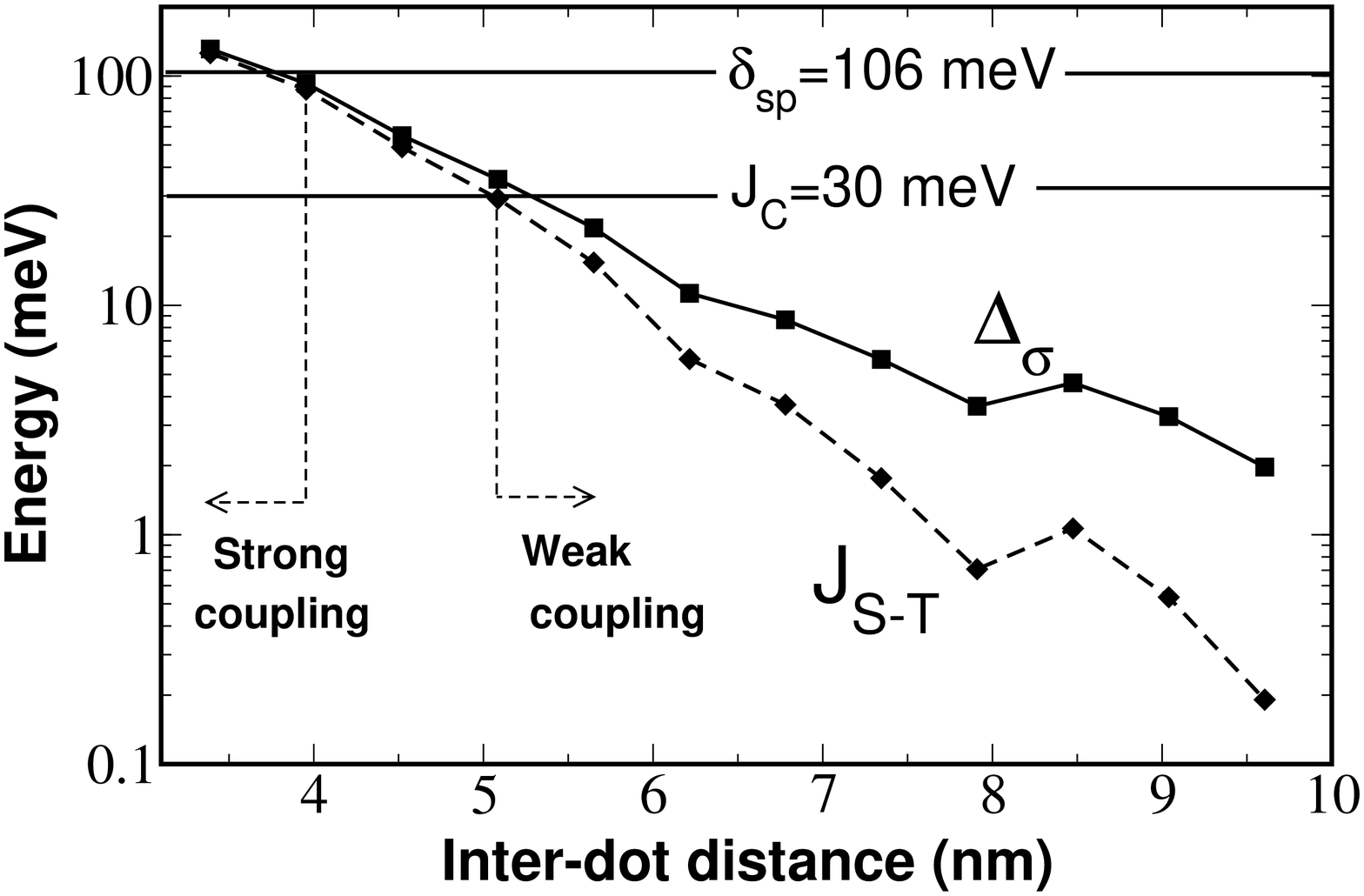}
\caption{The bonding-antibonding splitting
$\Delta_{\sigma}=\epsilon(\sigma_g)-\epsilon(\sigma_u)$ 
(solid line) and 
singlet-triplet splitting $J_{S-T}=E(^3\Sigma)-E(^1\Sigma_g^{(a)})$ 
(dashed line) {\it vs} inter-dot distance $d$.
We also show the single-dot s, p orbitals splitting 
$\delta_{sp}=e_s-e_p$ and the s orbital Coulomb interaction $J_C$. 
We define ``strong coupling'' by  $\Delta_{\sigma} \sim \delta_{sp}$ ($<$ 4 nm)
and ``weak coupling'',  $\Delta_{\sigma}\ll \delta_{sp}$ ($>$ 5 nm). 
}
\label{fig:t-j}
\end{figure}

\begin{figure}
\includegraphics[width=3.5in,angle=0]{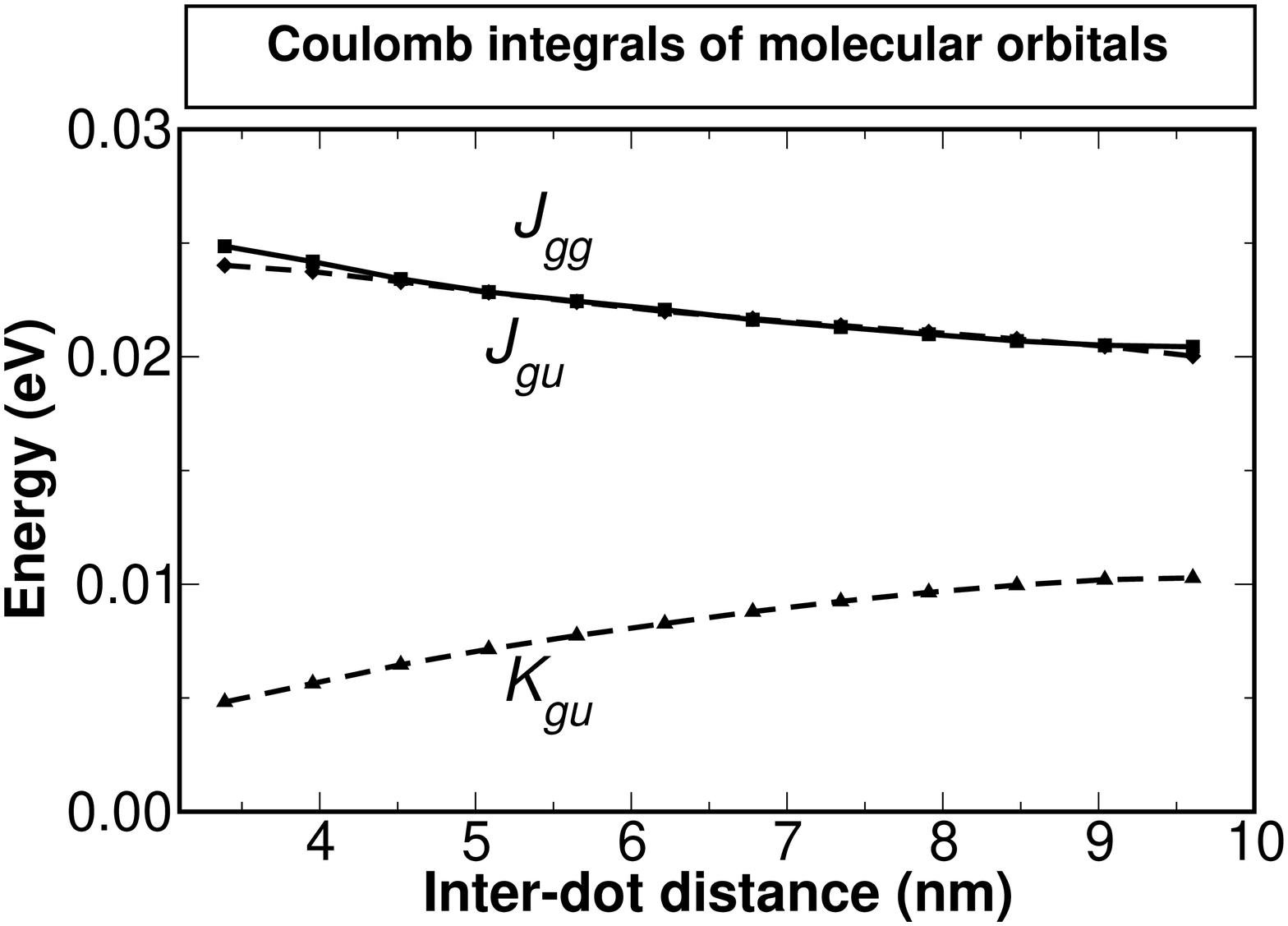}
\caption{Selected Coulomb integrals for molecular orbitals. $J_{gg}$ is the
 self-Coulomb energy of the
 $\sigma_g$ orbital and $J_{gu}$ is the Coulomb energy
 between $\sigma_g$ and $\sigma_u$ orbitals, while $K_{gu}$ is the exchange
 energy between $\sigma_g$ and $\sigma_u$ orbitals.
}
\label{fig:mo_col}
\end{figure}

We first study the two-electron states by solving the CI 
Eq.~(\ref{eq:coeff}), using all confined
molecular orbitals $\sigma_g$, $\sigma_u$ and $\pi_g$, $\pi_u$,
to construct the Slater determinants. This gives a 
total of 66 Slater determinants. 
The continuum states are far above the bound-state, and are thus
not included in the CI basis.
Figure~\ref{fig:mo_col} shows
some important matrix elements, including $J_{gg}$ (Coulomb energy of
$\sigma_g$ MO), $J_{gu}$ (Coulomb energy between $\sigma_g$ and
$\sigma_u$), and $K_{gu}$ (exchange energy between $\sigma_g$ and
$\sigma_u$). The Coulomb energy between $\sigma_u$ MO, $J_{uu}$ is nearly
identical to $J_{gg}$ and therefore is not plotted.
Diagonalizing the all-bound-state CI problem gives the 
two-particle states, shown in Fig.\ref{fig:states}a. 
We show all six $\Sigma$ states 
(where both electrons occupy the $\sigma$ states)
and the two lowest three-fold degenerate 
$^3\Pi_u$ states (where one electron occupies the 
$\sigma_g$ and one occupies one of the $\pi$ levels).
We observe that:

(a) The ground-state is singlet $^1\Sigma_g^{(a)}$ 
for all dot-dot distances.
However, the character of the state is very different at different 
inter-dot separation $d$, which can 
be analyzed by the isospin of the state,~\cite{palacios95} 
defined as the difference in the 
number of electrons occupying the
bonding ($N_B$) and antibonding ($N_{AB}$) states in a given CI state, i.e.
$I_z=(N_B-N_{AB})/2$, where $N_B$ and $N_{AB}$ are obtained from 
Eq.(\ref{eq:coeff}).
As shown in Fig.\ref{fig:Iz},
$I_z(d)$ of the $^1\Sigma_g^{(a)}$ state is very different
at different inter-dot distances:
At small inter-dot distance, the dominant configuration of the ground state is
$|\sigma_g^{\uparrow}, \sigma_g^{\downarrow\rangle}$ (both electrons occupy
bonding state and $N_B$=2), and $I_z$$\sim$1.
However, in the weak coupling region, there is significant mixing of
bonding $\sigma_g$ 
and anti-bonding states $\sigma_u$, and $I_z$ is smaller than 1, 
e.g $I_z$$\sim$ 0.2 at $d$= 9.5 nm. At
infinite separation, where the bonding and antibonding states 
are degenerate, one expects $I_z$$\rightarrow$ 0.  

(b) Next to the ground state, we find in Fig.\ref{fig:states}a 
the three-fold degenerate 
triplet states $^3\Sigma_u$, with $S_z$=1, -1 and 0.
In the absence of spin-orbit coupling, triplet states will not couple to
singlet states.
If we include spin-orbit coupling, the triplet may mix with the
singlet state, and the degeneracy will be lifted.
At large inter-dot distances, the ground state singlet $^1\Sigma_g^{(a)}$ and 
triplet states $^3\Sigma_u$ are degenerate.
The splitting of total CI energy between ground state singlet and triplet 
$J_{S-T}=E(^3\Sigma) - E(^1\Sigma_g)$ 
is plotted in Fig.\ref{fig:t-j} on a logarithmic scale.
As we can see, $J_{S-T}$ also decays approximately
exponentially between 4 and 8 nm,
and can be fitted as $J_{S-T}=5.28 \exp(-d/0.965)$ eV.
The decay length of 0.965 nm is shorter than the decay length 1.15 nm of
$\Delta_{\sigma}$. 
At small inter-dot separations, $J_{S-T} \sim
\Delta_{\sigma}$ in Fig.\ref{fig:t-j}, as expected from a simple 
Heitler-London model.~\cite{burkard99}  

(c) The two excited singlet states originating from the occupation of
$\sigma_u$ anti-bonding states,
$^1\Sigma_u$ and $^1\Sigma_g^{(b)}$ are further
above the $^3\Sigma_u$ state.
 
(d) The lowest $^3\Pi_u$ states are all triplet states. 
They are energetically very close to each other since we have two nearly
degenerate $\pi_u$ MO states. In the weak coupling region, 
the $\Pi_u$ states are well above the $\Sigma$ states, 
as a consequence of large single-particle 
energy difference $\epsilon(\pi_u) -\epsilon(\sigma_u)$.
However, the $\Pi_u$, and $^1\Sigma_g^{(b)}$ cross at about 4.5 nm,
where the single-particle MO level 
$\pi_u$ is still much higher than $\sigma_u$. In this case, the Coulomb
correlations have to be taken into account. 

In the following sections, we enquire as to possible, popularly
practiced simplifications 
over the all-bound-states CI treatment.

\subsection{Level-2 theory: reduced CI in the molecular basis}

In Level-2 theory, we will reduce the full 66$\times$66 CI problem of Level-1
to one that
includes only the $\sigma_g$ and $\sigma_u$ basis, giving a 6$\times$6 CI
problem.
The six many-body basis states
are shown in Fig.\ref{fig:CI_basis}a,
$|a\rangle$=$|\sigma_g^\uparrow,\sigma_u^\uparrow\rangle$, 
$|b\rangle$=$|\sigma_g^\downarrow,\sigma_u^\downarrow\rangle$,
$|c\rangle$=$|\sigma_g^\uparrow,\sigma_u^\downarrow\rangle$, 
$|d\rangle$=$|\sigma_g^\downarrow,\sigma_u^\uparrow\rangle$,
$|e\rangle$=$|\sigma_g^\uparrow,\sigma_g^\downarrow\rangle$, 
$|f\rangle$=$|\sigma_u^\uparrow,\sigma_u^\downarrow\rangle$.
In this basis set, the CI problem 
is reduced to a 6$\times$6 matrix eigenvalue equation, 
\begin{widetext}
\begin{equation}
\scriptsize
H= \left(\begin{array}{cccccc}
\epsilon_g+\epsilon_u+J_{gu}-K_{gu} & 0 & 0 &0 &0 &0 \\
0 & \epsilon_g+\epsilon_u+J_{gu}-K_{gu} & 0 & 0 &0 &0\\
0 & 0 & \epsilon_g+\epsilon_u+J_{gu}& -K_{gu}  
&-\Gamma^{gu}_{gg}  & -\Gamma^{gu}_{uu}\\
0& 0 &-K_{gu}&  \epsilon_g+\epsilon_u+J_{gu} &\Gamma^{gu}_{gg}   
&\Gamma^{gu}_{uu} \\
0& 0 &-\Gamma^{gg}_{gu}   & \Gamma^{gg}_{gu}  & 2\epsilon_g+J_{gg}   
& \Gamma^{gg}_{uu}\\
0& 0& -\Gamma^{uu}_{gu}   & \Gamma^{uu}_{gu}  
& \Gamma^{uu}_{gg} & 2\epsilon_u+J_{uu}
\end{array}\right)
\label{eq:morbit}
\normalsize
\end{equation}  
\end{widetext}
where, $\epsilon_g$ and $\epsilon_u$ are the single-particle energy levels 
for the MO's $|\sigma_g\rangle$ and $|\sigma_u\rangle$,
respectively. 
In the absence of spin-orbit coupling, 
the triplet states $|a\rangle$ and $|b\rangle$ are not 
coupled to {\it any} other 
states, as required by the total spin conservation, 
and thus they are already eigenstates. The rest of the
matrix can be solved using the integrals calculated from 
Eq.(\ref{eq:int_m}).
The results of the 6$\times$6 problem were compared (not shown)
to the all-bound-state
CI results: We find that 
the $\Sigma$ states of
Level-2 theory are very close to those of
the all-bound-state CI calculations, indicating a small coupling between 
$\sigma$ and $\pi$ orbitals in the {\it strong confinement} region.
We thus do not show graphically the results of Level-2.
However, since we use only $\sigma$ orbitals, the $\Pi$ states 
of Level-1 (Fig.\ref{fig:states}a)
are absent in Level-2 theory.
Especially, the important feature of crossover between $\Sigma$ 
and $\Pi_u$ states 
at 4 and 4.5 nm is missing. 
 
\subsection{Level-3 theory: single-configuration in the molecular basis}

\begin{figure}
\includegraphics[width=3.5in,angle=0]{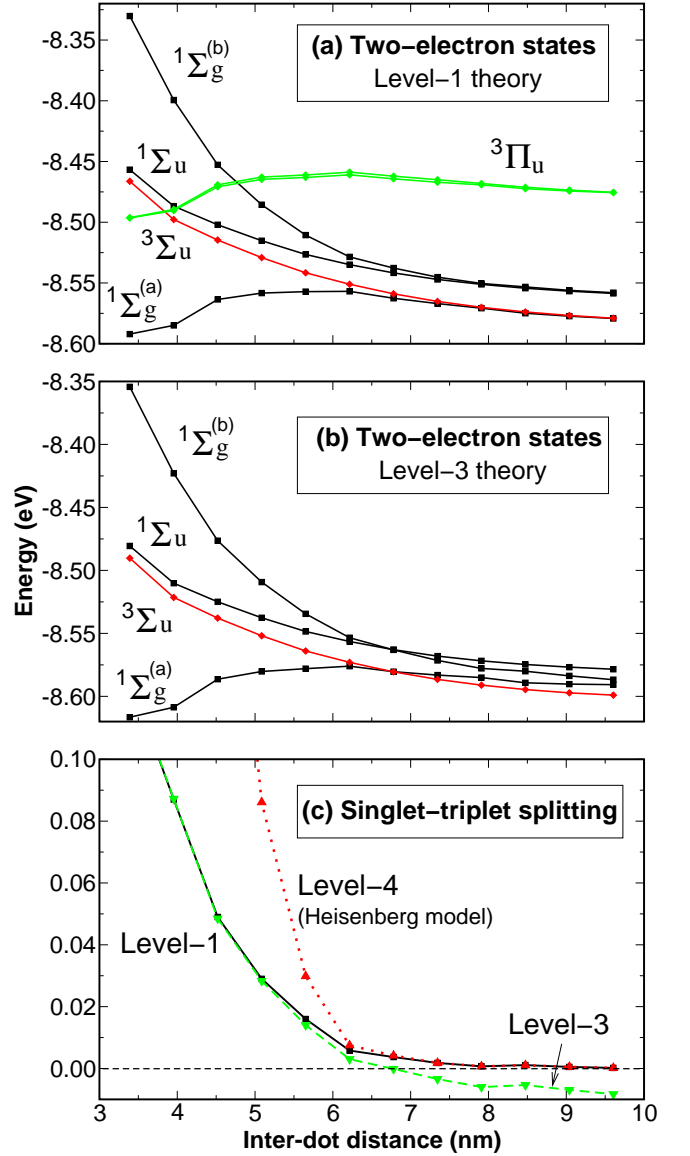}
\caption{ (Color online)
(a) Two-electron states calculated from CI using all confined MO from LCBB
(Level-1),
including the singlet $^1\Sigma_g^{(a)}$, $^1\Sigma_u$,
$^1\Sigma_g^{(b)}$ states and the 3-fold degenerated
triplet states $^3\Sigma_u$ as well as two 3-fold degenerated
triplet states $^3\Pi_u$.
(b) Two electron states calculated from the 
single-configuration approximation (Level-3).
(c) Comparison of the singlet-triplet splitting calculated from
Levels 1, 3 and 4 theories.
}
\label{fig:states}
\end{figure}

\begin{figure}
\includegraphics[width=3.5in,angle=0]{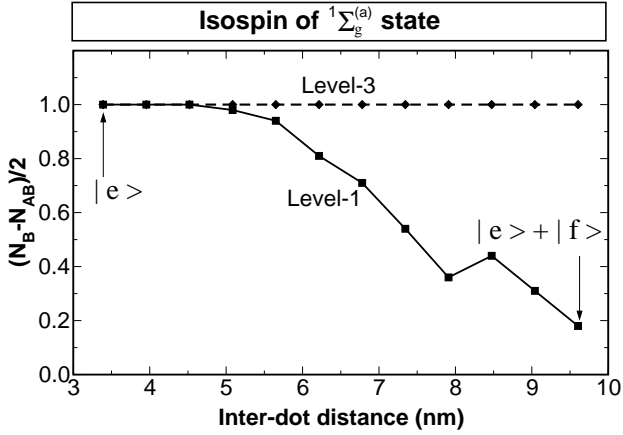}
\caption{Isospin, defined as the difference in the number of electrons
occupying the bonding ($N_B$) and antibonding ($N_{AB}$) states, 
of the $^1\Sigma_g^{(a)}$ state in Level-1 and Level-3
theories.
}
\label{fig:Iz}
\end{figure}

As is well known, mean-field-like treatments
such as RHF and LSD usually give incorrect dissociation behavior of molecules,
as the correlation effects are not adequately treated. 
Given that RHF and LSD are widely used in studying QMDs,
\cite{tamura98,nagaraja99,partoens00} 
it is important to
understand under which circumstance the methods will succeed and under which 
circumstance they will fail in describing the few-electron states in a
QDM.
In level-3 theory, we thus mimic the mean-field theory by 
further ignoring the {\it off-diagonal} Coulomb integrals in
Eq.(\ref{eq:morbit}) of Level-2 theory, 
i.e., we assume 
$\Gamma^{gu}_{uu}$= $\Gamma^{gu}_{gg}$
=$\Gamma^{uu}_{gg}$=0. 
This approximation is equivalent to ignoring
the coupling between
different configurations, and is thus called 
``single-configuration'' (SC) 
approximation.
At the SC level,  we have very simple analytical solutions
of the two-electron states,  
%
%\begin{widetext}
\begin{eqnarray}
E(^1\Sigma_g^{(a)}) &=&2\epsilon_g+J_{gg}; 
\quad\quad \quad\quad \quad\quad \;\;
 |^1\Sigma_g^{(a)}\rangle= |e\rangle\;\; , 
\label{eq:level3-Sga}\\
E(^3\Sigma_u)&=&(\epsilon_g +\epsilon_u) +J_{gu}-K_{gu}; \;
\left\{ \begin{array}{l} |^3_{+}\Sigma_u\rangle= |a\rangle\;, \\
                        |^3_{-}\Sigma_u\rangle= |b\rangle \;, \\
                        |^3_{0}\Sigma_u\rangle= |c\rangle-|d\rangle \; ,
\end{array} \right. 
\label{eq:level3-S3}
\\
E(^1\Sigma_u) &=&(\epsilon_g+\epsilon_u)  +J_{gu}+K_{gu} ;  \quad\;
 |^1\Sigma_u\rangle = |c\rangle+|d\rangle \;\; , 
\label{eq:level3-Su}
\\
E(^1\Sigma_g^{(b)})&=&2\epsilon_u +J_{uu} ;  
\quad\quad \quad\quad \quad\quad \;\;
 |^1\Sigma_g^{(b)}\rangle = |f\rangle \;\; .
\label{eq:level3-Sgb}
\end{eqnarray}
%\end{widetext}
%
The energies are plotted in Fig.\ref{fig:states}b.
When comparing the $\Sigma$ states of the SC approach
to the all-bound-state CI results
in Fig.\ref{fig:states}a, we find good agreement 
in the strong coupling region for $d\leq$ 5 nm (see Fig.\ref{fig:t-j}). 
However, the SC approximation fails qualitatively at larger 
inter-dot separations in two aspects:  
(i) The order of singlet state $^1\Sigma_g^{(a)}$ and triplet
state $^3\Sigma_u$ is reversed (see Fig.~\ref{fig:states}b,c). 
(ii) The $^1\Sigma_g^{(a)}$ and $^3\Sigma_u$ states fail
to be degenerate at large
interdot separation. 
This lack of degeneracy is also observed for
$^1\Sigma_g^{(b)}$ and $^1\Sigma_u$. 
These failures are due to the absence of correlations in the SC approximation.
Indeed as shown in Fig.\ref{fig:Iz},
the accurate Level-1 ground state singlet 
has considerable mixing of anti-bonding states, i.e. $I_z \rightarrow 0$ 
at large $d$. 
However, in the SC approximation
both electrons are {\it forced} to occupy the $\sigma_g$ orbital
in the lowest singlet state $^1\Sigma_g^{(a)}$ as a consequence of
the lack of the coupling between the configuration $|e\rangle$ of
Fig.\ref{fig:CI_basis}a and other
configurations. As a result, in Level-3 theory,
the isospins are forced to be $I_z$=1 for $^1\Sigma_g^{(a)}$ at 
all inter-dot distances $d$, 
which pushes the singlet energy higher than the triplet. 

\subsection{Level-4 theory: Hubbard model and Heisenberg model 
in a dot-centered basis}

The Hubbard model and the Heisenberg model are often used \cite{loss98} to
analyze entanglement and gate operations for
two spins qbits in a QDM.
Here, we analyze the extent to which such approaches can correctly capture the
qualitative physics given by more sophisticated models.
Furthermore, by doing so, 
we obtain the parameters of the models from realistic calculations.

\subsubsection{Transforming the states to a dot-centered basis}
Unlike the Level 1 - 3 theories, the
Hubbard and the Heisenberg models are written in a dot-centered basis
as shown in Fig.\ref{fig:CI_basis}b, rather than in the 
molecular basis of Fig.\ref{fig:CI_basis}a.
In a dot-centered basis, 
the Hamiltonian of Eq.(\ref{eq:morbit_h}) can be rewritten as,
\begin{widetext}
\begin{equation}
H=\sum_{\eta_1,\eta_2}
\sum_{\sigma} (e_{\eta_1} \delta_{\eta_1\eta_2}   +t_{\eta_1\eta_2})
\chi^{\dag}_{\eta_1,\sigma}\, \chi_{\eta_2\sigma}
+ {1\over2} \sum_{\eta_1, \cdots, \eta_4}
\sum_{\sigma,\sigma'}
\widetilde{\Gamma}^{\eta_1,\eta_2}_{\eta_3,\eta_4}\; 
\chi^{\dag}_{\eta_1,\sigma}\, \chi^{\dag}_{\eta_2,\sigma'}\,
\chi_{\eta_3,\sigma'}\,\chi_{\eta_4,\sigma}\, ,
\label{eq:ham_loc}
\end{equation}
\end{widetext}
where, $\eta=(l,p)$ and 
$\chi^{\dag}_{\eta,\sigma}$ creates an electron in 
the $l$=(s, p, $\cdots$) orbital on the $p$=(T, B) dot
with spin $\sigma$ that has single-particle energy $e_{\eta}$.
Here, $t_{\eta_1\eta_2}$ 
is the coupling between the $\eta_1$ and $\eta_2$ orbitals, and 
$\widetilde{\Gamma}^{\eta_1,\eta_2}_{\eta_3,\eta_4}$ 
is the Coulomb integral of single-dot orbitals
$\chi_{\eta_1}$, $\chi_{\eta_2}$, $\chi_{\eta_3}$ and $\chi_{\eta_4}$.

\begin{figure}
\includegraphics[width=3.5in,angle=0]{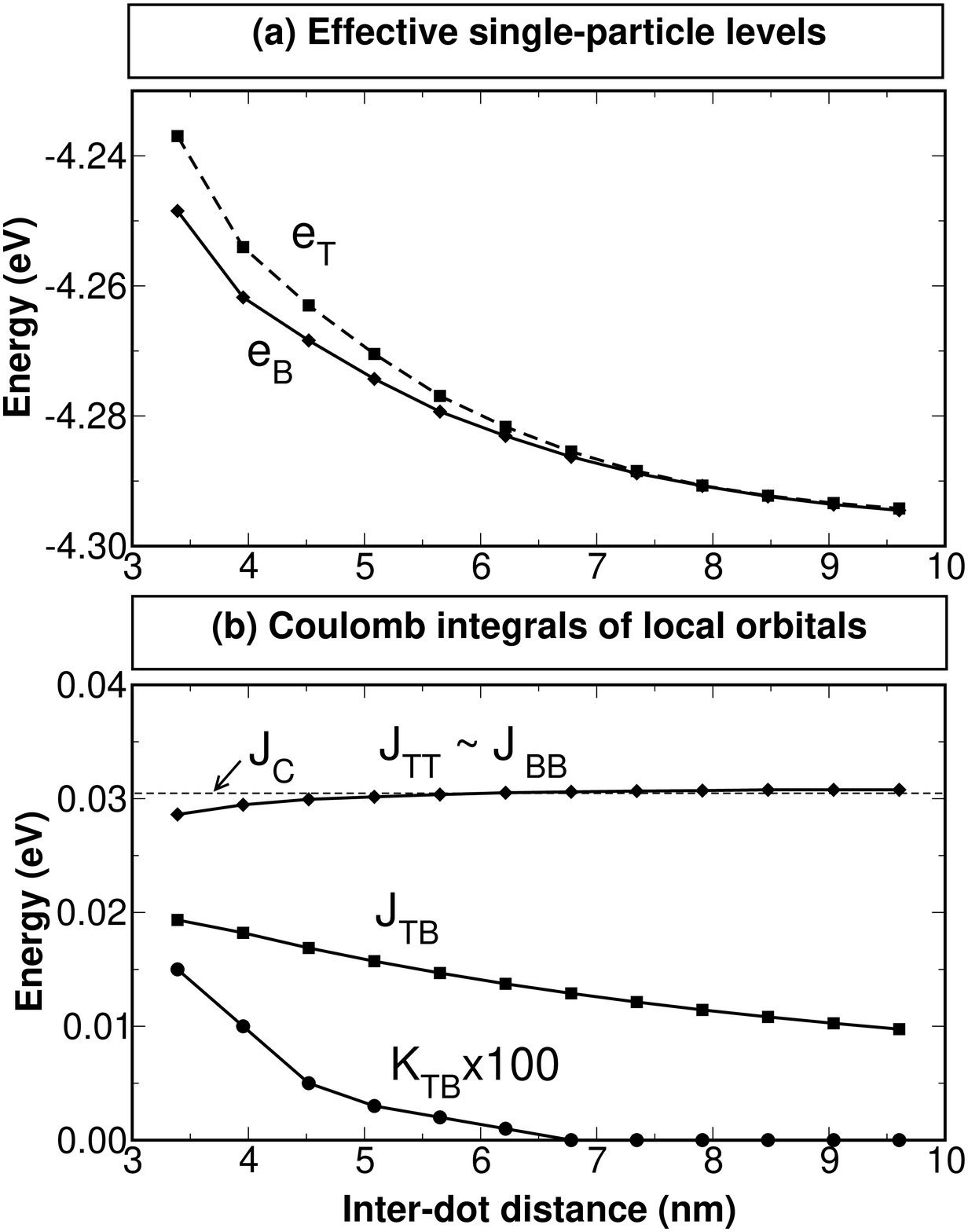}
\caption{
(a) Effective single-particle energy levels of s orbitals
localized on the top ($e_T$) and bottom ($e_B$) dots.
(b) Intra-dot Coulomb energy $J_{\rm TT}$, $J_{\rm BB}$,
inter-dot Coulomb energy
$J_{\rm TB}$ and inter-dot exchange energy $K_{\rm TB}$ (magnified by a factor
100). The dashed line
gives the single-dot s orbital self-Coulomb energy $J_C$.
}
\label{fig:loc-par}
\end{figure}

We wish to construct a Hubbard Hamiltonian whose 
parameters are taken from the fully atomistic single-particle theory.
To obtain such parameters in Eq.(\ref{eq:ham_loc}) including 
$e_{\eta}$, $t_{\eta_1\eta_2}$ 
and $\widetilde{\Gamma}^{\eta_1,\eta_2}_{\eta_3,\eta_4}$, 
we resort to a Wannier-like transformation, 
which transform the ``molecular''
orbitals (Fig.\ref{fig:CI_basis}a)
into single-dot ``atomic'' orbitals (Fig.\ref{fig:CI_basis}b).
The latter dot-centered orbitals
are obtained from a unitary rotation of the
{\it molecular} orbitals $\psi_{i}$, i.e, 
\begin{equation}
\chi_{\eta} =\sum_{i=1} \mathcal{U}_{\eta\, , i}\, \psi_i \; ,
\end{equation}
where, $\psi_{i}$ is the $i$-th molecular orbitals, $\chi_{\eta}$ is
the single dot-centered orbitals, and
$\mathcal{U}$ are unitary matrices, i.e. $\mathcal{U}^{\dag}\mathcal{U}=I$.
We chose the unitary matrices   
that maximize the total orbital 
self-Coulomb energy. The procedure of finding these  
unitary matrices is described in detail in Appendix~\ref{sec:append-a}.  
The dot-centered orbitals constructed this way
are approximately invariant to the change of
coupling between the dots.\cite{edmiston63}
Once we have the $\mathcal{U}$ matrices, we can obtain all the parameters in
Eq.(\ref{eq:ham_loc}) by transforming them from the molecular basis. 
The Coulomb integrals in the new basis set are given by Eq. (\ref{eq:col-int}),
while other quantities
including the effective single-particle
levels $e_{\eta}$ for the $\eta$-th dot-centered orbital, 
and the coupling between the $\eta_1$-th and $\eta_2$-th orbitals
$t_{\eta_1\eta_2}$
can be obtained from,
\begin{eqnarray}
e_{\eta}&=& 
\langle \chi_{\eta} | \hat{T}|\chi_{\eta} \rangle 
=\sum_{i} \mathcal{U}^*_{\eta\, ,i}\, \mathcal{U}_{\eta\, ,i}\, 
\epsilon_{i} \; ,
\label{eq:epsilon}\\
t_{\eta_1\eta_2} &=& \langle \chi_{\eta_1} | \hat{T}|\chi_{\eta_2} \rangle
 =\sum_{i} \mathcal{U}^*_{\eta_1,\, i}\, U_{\eta_2,\, i}\, 
\epsilon_i \; , 
\label{eq:t}
\end{eqnarray}
where $\epsilon_{i}$
is the single-particle level of the $i$-th {\it molecular} orbital,
and $\hat{T}$ is kinetic energy operator.  
Using the transformation of Eq.(\ref{eq:epsilon}), Eq.(\ref{eq:t})
and Eq.(\ref{eq:col-int}), we
calculate all parameters of Eq.(\ref{eq:ham_loc}).
Figure~\ref{fig:loc-par}a,
shows the effective single-dot energy 
of the ``s'' orbitals obtained in the Wannier representation for both top
and bottom dots. We see that
the effective single-dot energy levels 
increase rapidly for small $d$.  Furthermore the
energy levels for the top and bottom
orbitals are split due to the strain asymmetry between the two dots. 
We compute the Coulomb energies $J_{\rm TT}$, $J_{\rm BB}$ 
of the ``s'' orbitals on both top and
bottom dots, and the inter-dot Coulomb and exchange energies
$J_{\rm TB}$ and $K_{\rm TB}$ and  plot these 
quantities in Fig.~\ref{fig:loc-par}b.
Since $J_{\rm TT}$ and $J_{\rm BB}$ are very similar, 
we plot only $J_{\rm TT}$.
As we can see, the Coulomb energies of the dot-centered orbitals 
are very close to the Coulomb energy of the s orbitals 
of a {\it isolated} single dot (dashed line). 
The inter-dot Coulomb energy $J_{\rm TB}$ has comparable amplitude
to $J_{\rm TT}$ and decays slowly with distance, and remain very
significant, even at large separations. However, the exchange energy
between the orbitals localized on 
top dot and bottom dot $K_{\rm TB}$
is extremely small even when the dots are very close.

\subsubsection{``First-principles''
Hubbard model and Heisenberg model: Level-4}
In level-4 approximation, we use only the ``s'' orbital in each dot.
Figure~\ref{fig:CI_basis}b
shows all possible many-body basis functions of two electrons, where
top and bottom dots are denoted by 
``T'' and ``B'' respectively. 
The Hamiltonian in this basis set is, 
\begin{widetext}
\begin{equation}
\scriptsize
H= \left(\begin{array}{cccccc}
e_{\rm T}+e_{\rm B}+J_{\rm TB}-K_{\rm TB} 
&  0 &   0&   0  &0  & 0\\
0  & e_{\rm T}+e_{\rm B}+J_{\rm TB}
-K_{\rm TB} &  0&  0  &0 & 0 \\
0  &  0 &e_{\rm T}+e_{\rm B}
+J_{\rm TB} &-K_{\rm TB}&t-\widetilde{\Gamma}^{\rm TB}_{\rm BB} 
& t-\widetilde{\Gamma}^{\rm TB}_{\rm TT} \\
0  & 0   &-K_{\rm TB} &  e_{\rm T}
+e_{\rm B}+J_{\rm TB} &-t+\widetilde{\Gamma}^{\rm TB}_{\rm BB}  
& -t+\widetilde{\Gamma}^{\rm TB}_{\rm TT} \\
0  & 0   &  t-\widetilde{\Gamma}^{\rm BB}_{\rm TB}    
& -t+\widetilde{\Gamma}^{\rm BB}_{\rm TB}    
& 2e_{\rm B}+J_{\rm BB}  &0\\
0  & 0   &  t-\widetilde{\Gamma}^{\rm TT}_{\rm TB}    
& -t+\widetilde{\Gamma}^{\rm TT}_{\rm TB}   & 0    
& 2e_{\rm T} +J_{\rm TT}
\end{array}\right)\;.
\normalsize
\label{eq:hubbard}
\end{equation}
\end{widetext}
where $t=t_{TB}$ and to simplify the notation,
we ignore the orbital index ``s''.
If we keep all the matrix elements,
the description using the molecular basis of
Fig.~\ref{fig:CI_basis}a  
and the dot localized basis of Fig.~\ref{fig:CI_basis}b  
are equivalent, since they are connected by unitary
transformations. 
We now simplify Eq.(\ref{eq:hubbard}) by ignoring the
off-diagonal Coulomb integrals. 
The resulting Hamiltonian is the single-band Hubbard model.
Unlike Level-3 theory, in this case, 
ignoring off-diagonal Coulomb integrals (but keeping hopping)
can still give qualitatively correct results,
due to the fact that off-diagonal Coulomb integrals
such as  $\widetilde{\Gamma}^{\rm BB}_{\rm TB} \ll t$, and the correlation is
mainly carried by inter-dot hopping $t$.
We can further simplify the model by assuming
$e_{\rm T}=e_{\rm B}=\epsilon$;
$J_{\rm TT}=J_{\rm BB}=U$; and let $J_{\rm TB}=V$, $K_{\rm TB}=K$. 
We can then solve the simplified eigenvalue equation analytically.
The eigenvalues of the above Hamiltonian are (in order of increasing energy):\\
1. Ground state singlet $^1\Sigma_g^{(a)}$
\begin{equation}
E=2\epsilon + {1\over2} [ U+V+K - \sqrt{16 t^2 + (U-V-K)^2}]
\end{equation}
2. Triplet states (three-fold degenerate)  $^3\Sigma_u$
\begin{equation}
E=2\epsilon + V-K
\end{equation}
3. Singlet $^1\Sigma_u$
\begin{equation}
E=2\epsilon + U 
\end{equation}
4. Singlet $^1\Sigma_g^{(b)}$
\begin{equation}
E=2\epsilon + {1\over2} [ U+V+K + \sqrt{16 t^2 + (U-V-K)^2}] 
\end{equation}
In the Hubbard limit where Coulomb energy $U \gg t$, 
the singlet-triplet splitting 
$J_{S-T}=E(^3\Sigma)-E(^1\Sigma_g) \sim 4t^2/(U-V)$, 
which reduces the model to the
Heisenberg model
\begin{equation}
H= {4t^2\over U-V}\, \vec{S}_T\cdot\vec{S}_B\; ,
\end{equation} 
where $\vec{S}_T$ and $\vec{S}_B$ are the spin vectors on the top and bottom
dots. The Heisenberg model gives the
correct order for singlet and triplet states. 
The singlet-triplet splitting $J_{S-T}=4t^2/(U-V)$ 
is plotted in Fig.~\ref{fig:states}c and
compared to the results from all-bound-state
CI calculations (Level-1), and single-configuration approximations (Level-3). 
As we can see, at $d >$ 6.5 nm, 
the agreement between the Heisenberg model with CI is
good, but the Heisenberg model
greatly overestimates $J_{S-T}$ 
at $d<$ 6 nm.

\subsection{Comparison of 
pair correlation functions for Levels-1 to 4 theories}

\begin{figure}
\includegraphics[width=3.5in,angle=0]{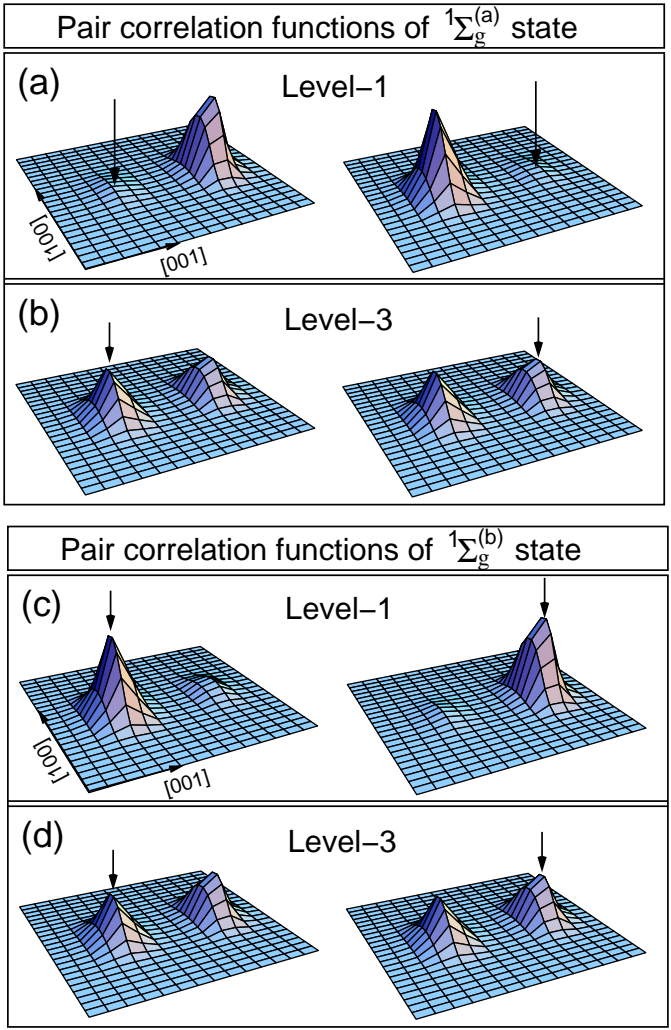}
\caption{(Color online) Comparison of pair correlation 
functions calculated from (a) Level-1, (b) Level-3 theory
for the $^1\Sigma_g^{(a)}$ state and (c) Level-1, (d) Level-3 theory
for the $^1\Sigma_g^{(b)}$ state at $d \sim$ 7 nm. 
On the left hand side, the first electron is fixed at the
center of the bottom dot, while on the right hand side, 
the first electon is fixed
at the center of the top dot, as indicated by the arrows.
}
\label{fig:corrfun}
\end{figure}

In the previous sections, we compared the energy levels of two-electron states
in several levels of approximations to all-bound-state CI results (Level-1).
We now provide further comparison of Levels 1-4 theories 
by analyzing the pair correlation functions 
and calculating the electron-electron entanglement at different levels of
approximations.

In Fig.~\ref{fig:corrfun} we show the pair correlation functions of 
Eq.(\ref{eq:corrfunc}) for the
$^1\Sigma_g^{(a)}$ and $^1\Sigma_g^{(b)}$
states at $d\sim$ 7 nm for Level-1 and Level-3 theories.
The correlation functions give the probability of finding the second electron
when the first electron is fixed at the position shown by the arrows at
the center of the bottom dot
(left hand side of Fig.~\ref{fig:corrfun}) 
or the top dot (right hand side of Fig.~\ref{fig:corrfun}).
Level-1 and Level-2 theories
give correlation-induced electron localization at large $d$: for
the $^1\Sigma_g^{(a)}$ state, the two electrons are localized on 
different dots,
while for the $^1\Sigma_g^{(b)}$ state, both electrons 
are localized on the same dot.\cite{he05b} 
In contrast, Level-3 theory shows delocalized states because of
the lack of configuration mixing. 
This problem is shared by RHF and LSD approximations.

\subsection{Comparison of 
the degree of entanglement for Levels 1-4 theories}

\begin{figure}
\includegraphics[width=3.5in,angle=0]{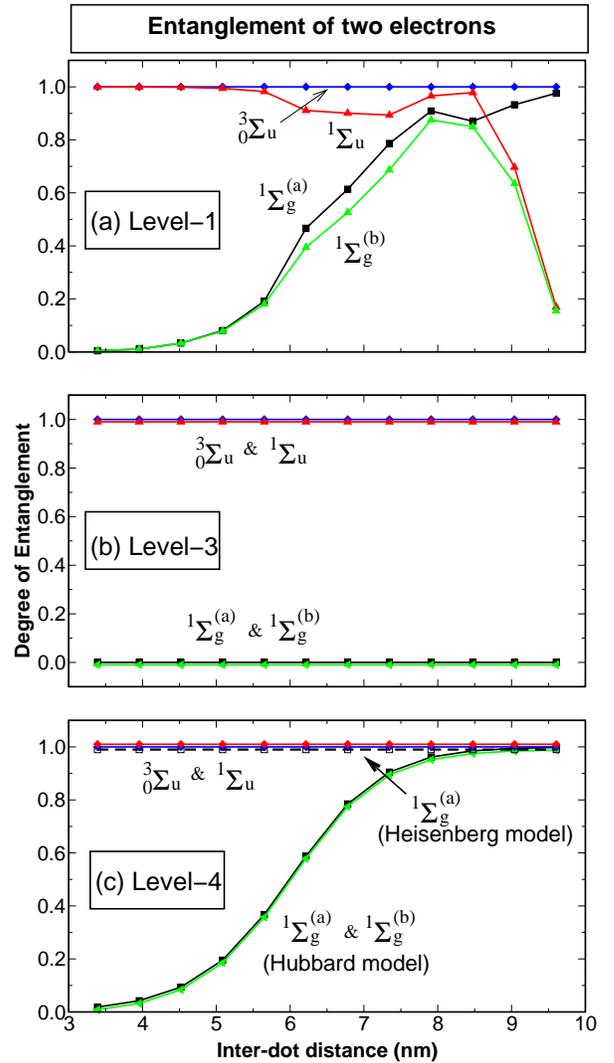}
\caption{(Color online) Comparison of the DOE calculated from (a) Level-1;
(b) Level-3 and (c) Level-4 theories for two-electron states. In panel (c),
both the DOE of the Hubbard model (solid lines) 
and of the Heisenberg model for $^1\Sigma_g^{(a)}$ state (dashed
line) are shown. 
}
\label{fig:DOE}
\end{figure}

The DOE of the four ``$\Sigma$'' states are plotted in Fig.\ref{fig:DOE} for 
Level-1, Level-3 and Level-4 theories; the DOE of 
Level-2 theory are virtually identical to those of Level-1 theory,
and are therefore not plotted.
We see that the Hubbard model
has generally reasonable agreement 
with Level-1 theory while the DOE calculated 
from Level-3 and Level-4 (Heisenberg model) theories
deviate significantly from the Level-1 theory, which is addressed below:

(i) {\it The $^1\Sigma_g^{(a)}$ state}: 
The Level-1 theory (Fig.\ref{fig:DOE}a), 
shows that the DOE of $^1\Sigma_g^{(a)}$ increases with $d$ and
approaches 1 at large $d$. The Hubbard model of Level-4 theory 
(Fig.\ref{fig:DOE}c) gives
qualitatively correct DOE for this state except for some details. 
However, Level-3 theory  (Fig.\ref{fig:DOE}b)
gives DOE $\mathcal{S}=0$ because the
wavefunction of $^1\Sigma_g^{(a)}$ is a single Slater determinant $|e\rangle$
[see Eq.(\ref{eq:level3-Sga})]. 
For the same reason, the DOEs of the $^1\Sigma_g^{(a)}$ state
in RHF and LSD approximations are also zero as a consequence of lack of
correlation. 
In contrast, the Heisenberg model of Level-4 theory gives  
$\mathcal{S}(^1\Sigma_g^{(a)})=1$.  This is because
the Heisenberg model assumes that the both
electrons are localized on different dots with zero double occupancy, and thus
overestimates the DOE.~\cite{Schliemann01,he05b}

(ii) {\it The $^1\Sigma_g^{(b)}$ state}: 
The Hubbard model gives the DOE 
of the $^1\Sigma_g^{(b)}$ state
identical to that of $^1\Sigma_g^{(a)}$ state. 
This is different from the result of
Level-1 theory, especially at large inter-dot separations. 
The difference comes from the
assumption in the Hubbard model that 
the energy levels and wavefunctions on the top
dot and on the bottom dot are identical
while as discussed in Ref.\onlinecite{he05b}, the wavefunctions 
are actually asymmetric
due to inhomogeneous strain in the real system. 
At $d >$ 8 nm, the $^1\Sigma_g^{(b)}$ state is the supposition of 
$|E\rangle$ and $|F\rangle$ configurations in the Hubbard model leading to
$\mathcal{S}=1$, 
while in Level-1 theory, the two electrons are both localized 
on the {\it top} dots ($|F\rangle$) at $d>$ 9 nm,\cite{he05b} 
resulting in near zero
entanglement.
For the same reason discussed in (i), the
Level-3 theory gives $\mathcal{S}(^1\Sigma_g^{(b)})=0$.

(iii) {\it The $^1\Sigma_u$ state}:  
Both the Level-3 theory and Hubbard model
give $\mathcal{S}(^1\Sigma_u)=1$. 
However, the $\mathcal{S}(^1\Sigma_u)$ of the Level-1 theory 
has more features as the consequence of the asymmetry of the system. 
In contrast to the $^1\Sigma_g^{(b)}$ state, in the
$^1\Sigma_u$ state, both electrons are 
localized on the {\it bottom} dot leading to near zero
entanglement at $d>$ 9 nm.

(iv) {\it The $^3\Sigma_u$ state}:
All levels of theories give very close results of DOE for  
$^3_0\Sigma_u$ state.
Actually, in Level-1 theory, the DOE of $^3_0\Sigma_u$ state 
is only slightly larger than 1, indicating 
weak entanglement of the
$\sigma$ and $\pi$ orbitals (the maximum entanglement one can get
from the total of six orbitals is $\mathcal{S}_{\rm max}={\rm log}_2\,6$), 
while in all other theories
(including the Level-2 theory) they are exactly 1 since 
these theories include only two $\sigma$ orbitals. 
The small coupling between $\sigma$ and $\pi$ orbitals 
is desirable for quantum computation, 
which requires the qbits states
to be decoupled from other states.

\section{summary}
\label{sec:summary}

We have shown the energy spectrum, pair-correlation functions
and degree of entanglement of two-electron states in  
self-assembled InAs/GaAs quantum
dot molecules {\it via} all-bound-state configuration interaction calculations 
and compared these quantities in different levels of approximations. 
We find that the correlation between electrons in the top and bottom dot is 
crucial to get the qualitative correct results for both the singlet-triplet
splitting and the 
degree of entanglement.
The single-configuration approximation and similar theories such as RHF,
LSD all suffer from lack of correlation and 
thus give incorrect ground state, singlet-triplet splitting $J_{S-T}$ and
degree of entanglement.
Highly simplified models, 
such as the Hubbard model gives qualitatively correct results for 
the ground state and $J_{S-T}$, 
while the Heisenberg model only give similar results at large $d$.
These two models are written in the dot-centered basis, where 
the correlation between the top and bottom dots are carried by the
single-particle tunneling.  
However, as a consequence of ignoring the asymmetry present
in the real system, 
the degree of entanglement calculated from the Hubbard model deviates
significantly from realistic atomic calculations. Moreover the
Heisenberg model greatly overestimates the degree of entanglement
of the ground state
as a consequence of further 
ignoring the electron double occupancy in the
dot molecule.

%%%%%%%%%%%%%%%%%%%%%%%%%%%%%%%%%%%%%%%%
\acknowledgments
%%%%%%%%%%%%%%%%%%%%%%%%%%%%%%%%%%%%%%%%

This work was founded by the U.S. Department of Energy, Office of Science,
  Basic Energy Science, Materials Sciences and Engineering, LAB-17 initiative,
under Contract No. DE-AC36-99GO10337 to NREL.

%%%%%%%%%%%%%%%%%%%%%%%%%%%%%%%%%%%%%%%%%%%%%%%%%%%%%%%%%%%%%%%%%%%%%%%%%%%%%%%

\appendix

\section{Degree of entanglement for two electrons}
\label{sec:append-b}

The entanglement is characterized
by the fact that the many-particle wavefunctions can not 
be factorized as a direct
product of single-particle wavefunctions. An entangled system displays 
non-locality which is one of the properties that distinguishes it 
from classic systems. 
So far, the only well established theory of entanglement pertains to 
two {\it distinguishable} particles,\cite{nielsen_book,bennett96b}
(e.g. electron and hole).
For a system made of two distinguishable particles $(A,B)$, 
the entanglement can be quantified by
von Neumann entropy of the partial density matrix of either $A$ or $B$,
\cite{nielsen_book,bennett96, wehrl78}
\begin{equation}
\mathcal{S}(A,B)
=-{\rm Tr}\; (\rho_A\; {\rm log}_2\, \rho_A) 
=-{\rm Tr}\; (\rho_B\; {\rm log}_2\, \rho_B)\; ,
\label{eq:von-entropy}
\end{equation}
where $\mathcal{S}(A,B)$ is the DOE of the state. 
$\rho_A$ and $\rho_B$ are the reduced density 
matrices for subsystems $A$ and $B$.
An alternative way to define the DOE for two distinguishable
particles is through a Schmidt decomposition, where
two-non-identical-particle wavefunctions are
written in an bi-orthogonal basis,
\begin{equation}
\Psi(A,B)=\sum_{i} \lambda_{i} \; |i_A\rangle\otimes | i_B\rangle\, ,
\label{eq:schmidt}
\end{equation}
with $\lambda_i \geq 0$ and $\sum_i \lambda_i^2=1$. 
The number of nonzero $\lambda_i$ is called the Schmidt rank.
For a pure state $\Psi(A,B)$
of the composite system $(A,B)$, we have,
\begin{eqnarray} 
\rho_A&=&\sum_i \lambda_i^2\; 
|i_A\rangle \langle i_A| \nonumber \\
\rho_B &=& \sum_i \lambda_i^2\; 
|i_B\rangle \langle i_B| \, .
\end{eqnarray}
It is easy to show from Eq.~\ref{eq:von-entropy} that the DOE 
for the two distinguishable particles is,
\begin{equation}
\mathcal{S}(A,B)=-\sum_i \lambda_i^2\; {\rm log}_2\, \lambda_i^2 \;. 
\label{eq:schmidt-entropy}
\end{equation}
We see from Eq.(\ref{eq:schmidt}) that 
when and only when the Schmidt rank equals 1, the 
two-particle wavefunction can be written as a direct product of
two single-particle wavefunctions. 
In this case, we have $\lambda=1$, and  
$\mathcal{S}(A,B)=0$ from Eq.(\ref{eq:schmidt-entropy}).

A direct generalization of DOE of
Eq.(\ref{eq:schmidt-entropy})
for two identical particles is problematic.
Indeed, there is no general way to 
define the subsystem $A$
and $B$ for two {\it identical} particles. 
More seriously, since two-particle wavefunctions for 
identical particles
are non-factorable due to their built-in symmetry, 
one may tend to believe that all two identical fermions (or Bosons) are
in entangled Bell state.\cite{nielsen_book} 
However, inconsistency comes up in the limiting cases. 
For example, suppose that
two electrons are localized on each of the
two sites $A$ and $B$ that are far apart, where the two
electrons can be treated
as {\it distinguishable} particles by assigning $A$ and $B$ to each electron,
respectively.
A pure state $\Psi$ that has 
the spin up for $A$ electron and spin down for $B$ electron is
$\Psi({\bf x}_1, {\bf x}_2) 
=1/\sqrt{2}[\phi_{A\uparrow}({\bf x}_1)\phi_{B\downarrow}({\bf x}_2)
-\phi_{A\uparrow}({\bf x}_2)\phi_{B\downarrow}({\bf x}_1)]$.
At first sight, because of the anti-symmetrization, it would seem that
the two electron states can not be written as a direct product of two single
particle wavefunctions, so this state is maximally entangled.  
However, when the overlap between two wavefunctions is negligible, 
we can treat these two particles as if they were distinguishable particles
and ignore the anti-symmetrization without any physical effect,
i.e. $\Psi({\bf x}_1, {\bf x}_2) 
=\phi_{A\uparrow}({\bf x}_1)\phi_{B\downarrow}({\bf x}_2)$. 
In this case, apparently the two electrons are {\it unentangled}.   
More intriguingly, in quantum theory, {\it all} fermions have to 
be anti-symmetrized even for {\it non-identical} particles, which does not mean
they are entangled.   

To solve this obvious inconsistency, 
alternative measures of the DOE of two fermions have 
been proposed and discussed
recently,\cite{schliemann01a,paskauskas01,li01,zanardi02,
shi03,wiseman03,ghirardi04} but no general solution 
has been widely accepted as yet. 
Schliemann {\it et al} \cite{schliemann01a} 
proposed using Slater decomposition 
to characterize the
entanglement (or, the so called ``quantum correlation'' in 
Ref. \onlinecite{schliemann01a}) 
of two fermions as a counterpart of the 
Schmidt decomposition for distinguishable
particles.
Generally a two-particle wavefunction can be written as,
\begin{equation}
\Psi=\sum_{i,j} \omega_{ij} |i\rangle \otimes |j\rangle \; ,
\label{eq:wav}
\end{equation}
where $|i\rangle$, $|j\rangle$ are the single particle orbitals.
The coefficient $\omega_{ij}$ must be antisymmetric for two fermions.
It has been shown in Ref.~\onlinecite{yang62, schliemann01a} 
that one can do a Slater
decomposition 
of $\omega_{ij}$ similar to the Schmidt decomposition for two
non-identical particles. It has been shown that $\omega$ can be block
diagonalized  through a unitary rotation of the single particle states,
\cite{yang62, schliemann01a} i.e., 
\begin{equation}
\omega'= U\, \omega\, U^{\dag} = {\rm diag} [Z_1, Z_2, \cdots, Z_r, Z_0] \; ,
\end{equation} 
where, 
\begin{equation}
Z_i=\left(\begin{array}{cc}
0 & z_i \\
-z_i & 0
\end{array}\right)\;.
\label{eq:z_i}
\end{equation}
and $Z_0=0$. Furthermore, $\sum_{i} z_{i}^2$=1, 
and $z_i$ is a non-negative real number. 
A more concise way to write down the state $\Psi$ is to use
the second quantization representation, 
\begin{equation}
\Psi=\sum_{i} z_{i}\; f_{2i-1}^{\dag}f_{2i}^{\dag} |0\rangle \; ,
\label{eq:f-dec} 
\end{equation}
where,  $f_{2i-1}^{\dag}$ and $f_{2i}^{\dag}$ are the creation 
operators for modes $2i-1$ and $2i$. 
Following Ref.\onlinecite{yang62}, it is easy to prove
that $z_i^2$ are eigenvalues of $\omega^{\dag}\omega$. 
The number of non-zero $z_i$ is called Slater rank.\cite{schliemann01a}
It has been argued in Ref.\onlinecite{schliemann01a}
that if the wavefunction can be written as single
Slater determinant, i.e., the Slater rank equals 1, 
the so called quantum correlation of the state is zero. 
The quantum correlation function defined 
in Ref.\onlinecite{schliemann01a} has similar properties, but nevertheless is
inequivalent to the usual definition of DOE.

Here, we propose a generalization of the DOE of
Eq.(\ref{eq:schmidt-entropy})
to two fermions, using the
Slater decompositions, 
\begin{equation}
\mathcal{S}=-\sum_{i} z_{i}^2 \;{\rm log}_2\, z_{i}^2\;.
\label{eq:my-entropy}
\end{equation}
The DOE measure of Eq. (\ref{eq:my-entropy}) has the following properties:

(i) This DOE measure is similar to the one proposed by   
Pa\u{s}kauskas {\it et al}, \cite{paskauskas01} and Li {\it et al},\cite{li01}
except that a different normalization condition is used.
In our approach,
the state of Slater rank 1 is unentangled, i.e.,
$\mathcal{S}$=0. In contrast,
Pa\u{s}kauskas {\it et al}, \cite{paskauskas01} and Li {\it et al},\cite{li01}
concluded that the unentangled
state has $\mathcal{S}={\rm ln}\, 2$, which is contradictory to
the fact that for distinguishable particles, an unentangled
state must has $\mathcal{S}$=0.  
In our approach, the maximum entanglement that a state can have is 
$\mathcal{S}={\rm log}_2\,N$, where $N$ is the number of single 
particle states. 

(ii)The DOE measure of Eq. (\ref{eq:my-entropy})
is invariant under any unitary transition of 
the single particle orbitals. Suppose there is
coefficient matrix $\omega$, a unitary transformation
of the single particle basis leads to a new 
matrix $\omega'=U\, \omega\, U^{\dag}$ 
and $\omega'^{\dag}\omega' 
= U\, \omega^{\dag} \omega \, U^{\dag}$. Obviously, 
this transformation would not
change the eigenvalue of $\omega^{\dag} \omega$, i.e., would not change
the entanglement of the system. 

(iii) The DOE of Eq. (\ref{eq:my-entropy}) 
for two fermions reduces to usual DOE 
measure of Eq. (\ref{eq:schmidt-entropy}) 
for two {\it distinguishable} particles in the
cases of zero double occupation of same site
(while the DOE measure proposed by
Pa\u{s}kauskas {\it et al}, \cite{paskauskas01} 
and Li {\it et al},\cite{li01} does not).
This can be shown as follows:
since the DOE of measure Eq. (\ref{eq:my-entropy}) is
basis independent, we can choose dot-localized basis set, 
(which in the case here is the top (T) and bottom (B) dots
[Fig.\ref{fig:CI_basis}(b)]), such that 
the antisymmetric $\omega$ matrix in the dot-localized basis 
has four blocks, 
\begin{equation}
\omega = \left( \begin{array}{cc}  
\omega_{TT} & - \omega_{TB}^{\dag} \\
\omega_{TB} & \omega_{BB} \end{array} \right) \; ,
\end{equation}
where, $\omega_{TT}$ is the coefficient matrix of two electrons both 
occupying the top dot, etc. 
If the double occupation is zero, i.e., two electrons are always on 
different dots, we have
matrices $\omega_{TT}=\omega_{BB}$=0. 
It is easy to show that $\omega^{\dag}\,\omega$
has two identical sets of eigenvalues $z_i^2$, each are the
eigenvalues of $\omega_{BT}^{\dag}\,\omega_{BT}$. 
On the other hand, if we treat the two electrons as distinguishable particles, 
and ignore the anti-symmetrization in the two-particle wavefunctions,
we have $\rho_B=\omega_{TB}^{\dag}\, \omega_{TB}$ and 
$\rho_T=\omega_{BT}^{\dag}\, \omega_{BT}$.
It is straightforward to show that in this case Eq. (\ref{eq:my-entropy})
and Eq. (\ref{eq:schmidt-entropy}) are equivalent.

\section{Construction of dot-centered orbitals}
\label{sec:append-a}

When we solve the single-particle Eq.(\ref{eq:schrodinger}) for the QDM, 
we get a set of molecular orbitals. However sometimes we need to
discuss the physics in a dot-localized basis set. 
The dot-localized orbitals $\chi_{\eta}$
can be obtained from a unitary rotation of molecular orbitals, 
\begin{equation}
\chi_{\eta} =\sum^N_{i=1} \mathcal{U}_{\eta\, ,i}\; \psi_{i} \, ,
\end{equation}
where, $\psi_{i}$ is the $i$-th molecular orbital, and 
$\mathcal{U}$ is a unitary matrix, i.e. $\mathcal{U}^{\dag}\mathcal{U}=I$.
To obtain a set of well localized orbitals, 
we require that the unitary matrix $\mathcal{U}$  
maximizes the total orbital self-Coulomb energy. The orbitals fulfilling the
requirement are approximately invariant under the changes due to
coupling between the dots.~\cite{edmiston63}
For a given unitary matrix $\mathcal{U}$, the Coulomb integrals in the rotated
basis are,
\begin{equation}
\widetilde{\Gamma}^{\eta_1,\eta_2}_{\eta_3,\eta_4} 
=\sum_{i,j,k,l}
\mathcal{U}^*_{\eta_1\, ,i}\, \mathcal{U}^*_{\eta_2\,,j}\,
\mathcal{U}_{\eta_3\, ,k}\,\mathcal{U}_{\eta_4\, ,l}\; 
\Gamma^{i,j}_{k, l} \, ,
\label{eq:col-int}
\end{equation} 
where $\Gamma^{i,j}_{k, l}$ are the Coulomb integrals in the molecular basis.
Thus, the total self-Coulomb energy for the orbitals $\{\chi_{\eta}\}$ is:
\begin{equation}
U_{\rm tot}=\sum_{\eta}
\widetilde{\Gamma}^{\eta,\eta}_{\eta,\eta}
=\sum_{\eta} \sum_{i,j,k,l}
\mathcal{U}^*_{\eta\, ,i}\,\mathcal{U}^*_{\eta\, j}\,
\mathcal{U}_{\eta\, ,k}\, \mathcal{U}_{\eta \, , l}\;
\Gamma^{i, j}_{k, l} \, .
\end{equation}  
The procedure of finding the unitary matrix $\mathcal{U}$ that maximizes 
$U_{\rm tot}$ is similar to the procedure 
given in Ref. \onlinecite{marzari97}
where the  maximally localized Wannier functions for extended systems are
constructed using a different criteria.
Starting from $\mathcal{U}=I$, we find a new 
$\mathcal{U}=I+\delta\epsilon$ that increases $U_{\rm tot}$.
To keep the new matrix unitary, we require $\delta \epsilon$
to be a small anti-Hermitian matrix. It is easy to prove that 
\begin{equation}
G_{i, j}\equiv{\delta U_{\rm tot} \over \delta \epsilon_{j,i} } 
= \Gamma^{j,j}_{i,j} 
+\Gamma^{j,j}_{j,i}-\Gamma_{i,i}^{i,j} 
-\Gamma_{i,i}^{j,i}  
\end{equation}
and to verify that $G_{i,j}=-G^*_{j,i}$.
By choosing $\delta \epsilon_{i,j}=-\epsilon G_{i,j}$, 
where $\epsilon$ is a small real
number, we always have (to the first-order of approximation) 
$\Delta U_{\rm tot}= \epsilon \|G\| \ge0$, i.e. 
the procedure always increases the total self-Coulomb energy. 
To keep the strict unitary character of 
the $\mathcal{U}$ matrices in the procedure, 
the $\mathcal{U}$ matrices
are actually updated
as $\mathcal{U}\leftarrow\mathcal{U}\exp (-\epsilon G)$,
until the localization is achieved. 

%--------------------------------------------------------------------

%\bibliography{DotBiblio,books}% Produces the bibliography via BibTeX.

\end {document}